\documentclass[aps,prl,twocolumn,showpacs,floatfix,nobibnotes,superscriptaddress,amsmath,amssymb,longbibliography,preprintnumbers]{revtex4-2}

\usepackage{graphicx}
\usepackage{epsfig}
\usepackage{bm}
\usepackage{color}
\usepackage{float}
\usepackage{dcolumn}
\usepackage{multirow} 
\usepackage{hyperref}
\usepackage{mathtools}
\usepackage{xspace}
\usepackage{natbib}
\usepackage{bibentry}

\newcommand{\be}{\begin{equation}}
\newcommand{\ee}{  \end{equation}}
\newcommand{\ba}{\begin{eqnarray}}
\newcommand{\ea}{  \end{eqnarray}}
\newcommand{\bas}{\begin{eqnarray*}}
\newcommand{\eas}{  \end{eqnarray*}}

\begin{document}

\title{Coupled-cluster calculations of neutrinoless double-beta decay in $^{48}$Ca}
\preprint{LA-UR-21-22076}

\author{S.~J.~Novario}
\affiliation{Department of Physics and Astronomy, University of Tennessee,
  Knoxville, TN 37996, USA} 
\affiliation{Physics Division, Oak Ridge National Laboratory,
  Oak Ridge, TN 37831, USA} 

\author{P.~Gysbers}
\affiliation{TRIUMF, 4004 Wesbrook Mall, Vancouver BC, V6T 2A3, Canada}
\affiliation{Department of Physics and Astronomy, University of British Columbia, Vancouver BC, V6T 1Z1, Canada}

\author{J. Engel}
\affiliation{Department of Physics, University of North Carolina, Chapel Hill, NC 27514, USA}

\author{G.~Hagen}
\affiliation{Physics Division, Oak Ridge National Laboratory,
  Oak Ridge, TN 37831, USA} 
\affiliation{Department of Physics and Astronomy, University of Tennessee,
  Knoxville, TN 37996, USA} 
\affiliation{TRIUMF, 4004 Wesbrook Mall, Vancouver BC, V6T 2A3, Canada}

\author{G.~R.~Jansen}
\affiliation{National Center for Computational Sciences, Oak Ridge National
  Laboratory, Oak Ridge, TN 37831, USA}
\affiliation{Physics Division, Oak Ridge National
  Laboratory, Oak Ridge, TN 37831, USA}

\author{T.~D.~Morris}
\affiliation{Physics Division, Oak Ridge National Laboratory,
  Oak Ridge, TN 37831, USA} 

\author{P. Navr{\'a}til}
\affiliation{TRIUMF, 4004 Wesbrook Mall, Vancouver BC, V6T 2A3, Canada}

\author{T.~Papenbrock}
\affiliation{Department of Physics and Astronomy, University of Tennessee,
  Knoxville, TN 37996, USA}
\affiliation{Physics Division, Oak Ridge National Laboratory,
  Oak Ridge, TN 37831, USA}

\author{S.~Quaglioni}
\affiliation{Lawrence Livermore National Laboratory, P.O. Box 808, L-414,
  Livermore, California 94551, USA}


\begin{abstract}
  We use coupled-cluster theory and nuclear interactions from chiral
  effective field theory to compute the nuclear matrix element for the
  neutrinoless double-beta decay of $^{48}$Ca. Benchmarks with the
  no-core shell model in several light nuclei inform us about the
  accuracy of our approach. For $^{48}$Ca we find a relatively small
  matrix element. We also compute the nuclear matrix element for the
  two-neutrino double-beta decay of $^{48}$Ca with a quenching factor
  deduced from two-body currents in recent ab-initio calculation of
  the Ikeda sum-rule in $^{48}$Ca
  [Gysbers {\it et al.}, Nature Physics {\bf 15}, 428–431 (2019)].
\end{abstract}

\maketitle 

{\it Introduction and main result.--- } Neutrinoless double-beta
($0\nu\beta\beta$) decay is a hypothesized electroweak process in
which a nucleus undergoes two simultaneous beta decays but emits no
neutrinos~\cite{furry1939}.  The observation of this lepton-number
violating process would identify the neutrino as a Majorana particle
(i.e.\ as its own antiparticle)~\cite{schechter1982} and provide
insights into both the origin of neutrino
mass~\cite{minkowski1977,mohapatra1980} and the matter-antimatter
asymmetry in the universe~\cite{davidson2008}. Experimentalists are
working intently to observe the decay all over the world; current
lower limits on the lifetime are about
$10^{26}$~y~\cite{anton2019,alvis2019,agostini2019},
and sensitivity will be improved by two orders of magnitude in the
coming years.

Essential for planning and interpreting these experiments are nuclear
matrix elements (NMEs) that relate the decay lifetime to the Majorana
neutrino mass scale and other measures of lepton-number
violation. Unfortunately, these matrix elements are not well known and
cannot be measured. Computations based on different models and
techniques lead to numbers that differ by factors of three to five
(see Ref.~\cite{engel2017} for a recent review).  Compounding these
theoretical challenges is the recent discovery that, within chiral
effective field theory
(EFT)~\cite{vankolck1994,bedaque2002,epelbaum2009,machleidt2011}, the
standard long-range $0\nu\beta\beta$ decay operator must be
supplemented by an equally important zero-range (contact) operator of
unknown strength~\cite{cirigliano2018}.  Efforts to compute the
strengths of this contact term from quantum chromodynamics
(QCD)~\cite{cirigliano2020, cirigliano2020b} and attempts to better
understand its impact are underway~\cite{cirigliano2019}.

The task theorists face at present is to provide more accurate
computations of $0\nu\beta\beta$ NMEs, including those associated with
contact operators, and quantify their uncertainties. In this Letter,
we employ the coupled-cluster method to perform first-principle
computations of the matrix element that links the $0\nu\beta\beta$
lifetime of $^{48}$Ca with the Majorana neutrino mass scale.  Among
the dozen or so candidate nuclei for $0\nu\beta\beta$ decay
experiments ~\cite{barabash2015}, $^{48}$Ca stands out for its fairly
simple structure, making it amenable for an accurate description based
on chiral EFT and state-of-the-art many-body
methods~\cite{hagen2015}. By varying the details of our calculations,
we will estimate the uncertainty of our prediction. To gauge the
quality of our approach we also compute the two-neutrino double-beta
decay of $^{48}$Ca and compare with data. Our results will directly
inform $0\nu\beta\beta$ decay experiments that use
$^{48}$Ca~\cite{tetsuno2020} and serve as an important stepping stone
towards the accurate prediction of NMEs in $^{76}$Ge, $^{130}$Te, and
$^{136}$Xe, which are candidate isotopes of the next-generation
$0\nu\beta\beta$ decay experiments. Calculations in those
nuclei presumably require larger model spaces, inclusion of
tri-axial deformation, and symmetry projection.

\begin{figure}
  \includegraphics[width=0.5\textwidth]{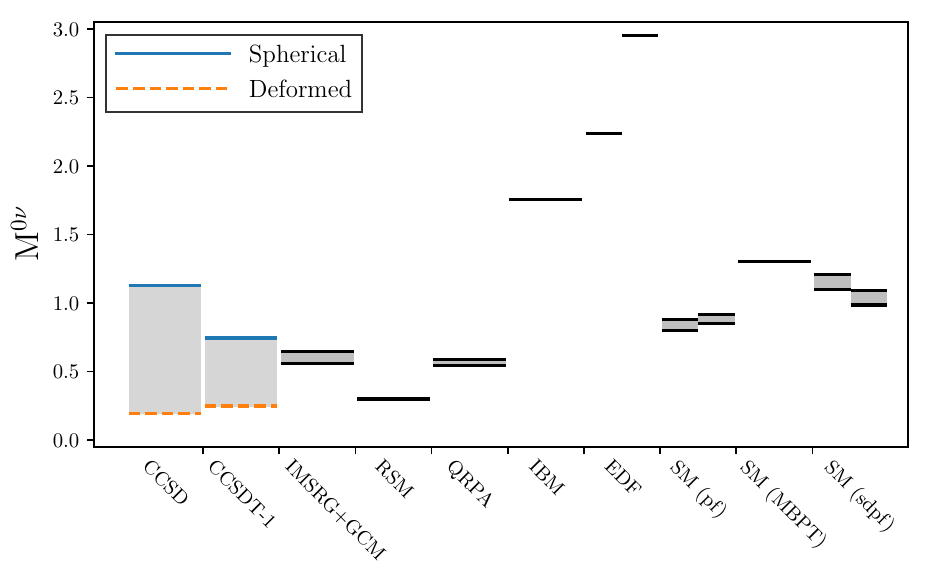}
  \caption{(Color online) Comparison of the NME for the
  $0\nu\beta\beta$ decay of $^{48}$Ca, calculated within various approaches (see
  text for details). The coupled-cluster results use both the CCSD and CCSDT-1
  approximations with both the spherical and deformed reference states. For
  IMSRG+GCM, the double bars show the effects of uncertainty in model-space
  size; otherwise they show those of uncertainty in short-range correlation
  functions.} \label{ca48-0vbb}
\end{figure}

Figure~\ref{ca48-0vbb} shows several recent results for the NME
governing the $0\nu\beta\beta$ decay $^{48}$Ca$\to^{48}$Ti and
compares them with those of this work. The coupled cluster results
obtained here, with both the CCSD and CCSDT-1 approximations
(explained below), display uncertainties from details of the
computational approach. They are compared to the very recent {\it ab
  initio} results from the in-medium similarity group renormalization
method with the generator coordinator method
(IMSRG+GCM)~\cite{yao2020}, a realistic shell-model
(RSM)~\cite{coraggio2020}, the quasi-particle random phase
approximation (QRPA)~\cite{simkovic2013}, the interacting boson model
(IBM)~\cite{barea2015}, various energy-density functionals
(EDF)~\cite{vaquero2013,yao2015}, and several more phenomenological
shell model (SM) calculations. The latter either limit themselves to
the $pf$-shell~\cite{senkov2013,menendez2009}, include perturbative
corrections from outside of the $pf$-shell~\cite{kwiatkowski2014}, or
are set in the $sdpf$ shell-model space~\cite{iwata2016}.  We see that
the {\it ab initio} results of this work and of Ref.~\cite{yao2020}
are consistent with each other and with the most recent
work~\cite{belley2020}.  Our result, in the CCSDT-1 approximation, is
$0.25 \le M^{0\nu} \le 0.75$.

{\it Method.--- } We employ the intrinsic Hamiltonian
\begin{equation}
  \label{intham}
  H = \sum_{i<j}\left({(\vec{p}_i-\vec{p}_j)^2\over 2mA} + V
    _{NN}^{(i,j)}\right) + \sum_{ i<j<k}V_{NNN}^{(i,j,k)} .
\end{equation}
Here $m$ is the nucleon mass, $\vec{p}$ is the momentum operator, $A$ is the
mass number of the nucleus, and $V_{NN}^{(i,j)}$ and $V_{NNN}^{(i,j,k)}$ are the
nucleon-nucleon (NN) and three-nucleon (NNN) potentials, respectively.  We
employ the chiral potential 1.8/2.0 (EM) of Ref.~\cite{hebeler2011}.
Three-nucleon force contributions are limited to those from matrix elements
in the oscillator basis with $N_1 + N_2 + N_3 \le 16$, where $N_i = 2n_i + l_i$
are single-particle energies. The oscillator basis has a frequency
$\hbar\Omega = 16\ \mathrm{MeV}$ and we find that working within a model space
with $N_i = 10$ is sufficient to produce converged results.

Following Refs.~\cite{tichai2019,novario2020}, we transform the
Hamiltonian from the spherical oscillator basis to a natural-orbital
basis by diagonalizing the one-body density matrix. We denote the
resulting reference state, i.e.\ the product state constructed from
the $A$ single-particle states with largest occupation numbers, by
$|\Phi_0\rangle$ and the Hamiltonian that is normal-ordered with
respect to this non-trivial vacuum by $H_N$.  We retain $NNN$ forces
at the normal-ordered two-body level~\cite{hagen2007a,roth2012}.

Coupled-cluster theory~\cite{coester1958,coester1960,cizek1966,
  cizek1969,kuemmel1978,bartlett2007,hagen2013c} is based on the
similarity-transformed Hamiltonian, $\overline{H}_N =
e^{-\hat{T}}{H_N}e^{\hat{T}}$.  The cluster operator $\hat{T}$ is a
sum of particle-hole (ph) excitations from the reference
$|\Phi_0\rangle$ and commonly truncated at the two-particle two-hole
($2p$--$2h$) or $3p$--$3h$ level. The amplitudes in $\hat{T}$ are
chosen so that the reference state $|\Phi_0\rangle$ becomes the right
ground state of $\overline{H}_N$.  Because $\overline{H}_N$ is
non-Hermitian, the left ground state is $\langle \Phi_0 \vert ( 1 +
\hat{\Lambda} )$, where $\hat{\Lambda}$ is a de-excitation operator
with respect to the reference~\cite{bartlett2007,hagen2013c}.  In this
paper, we work at the leading-order approximation to coupled-cluster
with singles-doubles-and-triples excitations (CCSDT), known as
CCSDT-1~\cite{watts1993,watts1995}. To make the computation feasible,
we truncate the $3p$--$3h$ amplitudes by imposing a cut on the product
of occupation probabilities $n_a$ for three particles above the Fermi
surface, $n_{a}n_{b}n_{c} \geq \mathcal{E}_{3}$, and for three holes
below the Fermi surface, $(1-n_{i})(1-n_{j})(1-n_{k}) \geq
\mathcal{E}_{3}$. This truncation favors orbitals near the Fermi
surface.  The limits are large enough so that all CCSDT-1 results
presented below are stable against changes in them.

We are interested in computing $\vert M^{0\nu} \vert^{2} =
\langle\Psi_\mathrm{I}\vert\hat{O}^\dagger_{0\nu}\vert\Psi_\mathrm{F}\rangle\langle\Psi_\mathrm{F}\vert\hat{O}_{0\nu}\vert\Psi_\mathrm{I}\rangle$,
where $\hat{O}_{0\nu}$ is the $0 \nu \beta \beta$ operator and
$\Psi_\mathrm{I}$ and $\Psi_\mathrm{F}$ denote the ground states of
the initial and final nuclei, respectively. Within coupled-cluster
theory, we can structure the calculation in two ways. In a first
approach, we can use the right and left ground states of $^{48}$Ca
($|\Phi_0\rangle$ and $\langle \Phi_0 \vert (1+\hat{\Lambda})$,
respectively) to compute
\begin{equation}
  \label{M0vbb_CC1}
  \vert M^{0\nu} \vert^{2} = \langle\Phi_0\vert (1+\hat{\Lambda})\overline{O^\dagger}_{0\nu} \hat{R}\vert \Phi_0 \rangle
  \langle \Phi_0 \vert \hat{L} \overline{O}_{0\nu} \vert \Phi_0 \rangle .
\end{equation}
In this case, we use equation-of-motion coupled-cluster (EOM-CC)
techniques~\cite{bartlett2007,shavittbartlett2009,jansen2011,jansen2012,hagen2012a,hagen2012b,binder2013a}
to represent the right and left $^{48}$Ti ground states (denoted by
$\hat{R} \vert \Phi_0 \rangle$ and $\langle\Phi_0 \vert \hat{L}$,
respectively) by generalized excited states of $^{48}$Ca with two more
protons and two less neutrons ~\cite{payne2019,liu2019}. Here, we also
work in the CCSDT-1 approximation. In Eq.~(\ref{M0vbb_CC1})
$\overline{O}_{0\nu} \equiv e^{-\hat{T}}\hat{O}_{0\nu}e^{\hat{T}}$ is
the similarity-transformed $0\nu\beta\beta$ operator.

In an alternative approach, we can decouple the ground state of the
final nucleus, i.e.\ take $|\Phi_0\rangle$ as a reference right ground
state for $^{48}$Ti (with $\langle \Phi_0 \vert (1+\hat{\Lambda})$ its
left ground state), and target the initial nucleus $^{48}$Ca with
EOM-CC. This procedure leads to the expression
\begin{equation}
  \label{M0vbb_CC2}
  \vert M^{0\nu} \vert^{2} = \langle \Phi_0 \vert \hat{L} \overline{O^\dagger}_{0\nu} \vert \Phi_0 \rangle
  \langle \Phi_0 \vert (1+\hat{\Lambda}) \overline{O}_{0\nu} \hat{R} \vert \Phi_0 \rangle,
\end{equation}
where the $^{48}$Ca right and left ground states
($\hat{R}\vert \Phi_0 \rangle$ and $\langle\Phi_0 \vert \hat{L}$,
respectively) are represented by generalized excited states of
$^{48}$Ti. Because the two approaches are identical only when the
cluster operators are not truncated, the difference between them is a
measure of the truncation effects. As the ground state of $^{48}$Ca is
spherical, the first procedure allows us to exploit rotational
symmetry. By contrast, starting from $^{48}$Ti introduces a deformed
(though axially symmetric) reference state, which accurately reflects
the non-trivial vacuum properties and captures static correlations
that would be many-particle--many-hole excitations in the spherical
scheme~\cite{ringschuck}. It comes at the expense of breaking
rotational invariance, which eventually could be restored with
symmetry restoration techniques \cite{duguet2015,henderson2017,tsuchimochi2018}.

In chiral EFT, the $0\nu\beta\beta$ operator is organized into a
systematically improvable expansion similarly to the nuclear
forces~\cite{cirigliano2018b}.  The lowest-order contributions to the
$0\nu\beta\beta$ operator are a long-range Majorana neutrino potential
that can be divided into three components, Gamow-Teller (GT), Fermi
(F), and tensor (T), that contain different combinations of spin
operators, with $\hat{O}_{0\nu} = \hat{O}^{\mathrm{GT}}_{0\nu} +
\hat{O}^{\mathrm{F}}_{0\nu} + \hat{O}^{\mathrm{T}}_{0\nu}$. The
corresponding two-body matrix elements, as is conventional, are taken
from Ref.~\cite{simkovic2008}, which adds form factors to the leading
and next-to-leading operators. We use the closure approximation (which
is sufficiently accurate~\cite{senkov2013}), with closure energies
$E_{\mathrm{cl}}=5$~MeV for all benchmarks in light nuclei and
$7.72$~MeV for the decay $^{48}$Ca$\to^{48}$Ti.

The NME for the $2\nu\beta\beta$ is similar to the $0\nu\beta\beta$ case
except the two-body operator is replaced by a double application of the
one-body Gamow-Teller operator, $\mathbf{\sigma}\tau^{-}$
\footnote{Here $\tau^{-}$ changes a neutron into a proton.}, with an
explicit summation over the intermediate $1^{+}$ states between them,
\begin{equation}
  \label{M2vbb_sum}
  \vert M^{2\nu} \vert^{2} = \left\vert\sum_{\mu}\frac{\langle 0^{+}_{F}\vert\mathbf{\sigma}\tau^{-}\vert 1^{+}_{\mu}\rangle\langle 1^{+}_{\mu}\vert\mathbf{\sigma}\tau^{-}\vert 0^{+}_{I}\rangle}{\Delta E_{\mu}+(E_{\mathrm{I}}-E_{\mathrm{F}})/2}\right\vert^{2}.
\end{equation}
The denominator consists of the excitation energy of the intermediate
states with respect to the initial ground state,
$\Delta E_{\mu}=E_{\mu}-E_{\mathrm{I}}$, and the energy difference
between the initial and final states, $E_{\mathrm{I}}-E_{\mathrm{F}}$
(see Supplemental Material and~\cite{vogel2012,kotila2012} for more details).
The direct computation of the matrix element~(\ref{M2vbb_sum}) would
require several tens of states in the intermediate nucleus and several
hundred Lanczos iterations, making it unfeasible in our large model space.

We note that the Green's function at the center of
this matrix element can be computed efficiently using the Lanczos
(continued fraction) method starting from a $1^+$ pivot
state~\cite{engel1992,haxton2005,marchisio2003,miorelli2016,rotureau2017}. We
generate Lanczos coefficients ($a_{i}, b_{i}$ and $a^{*}_{i},
b^{*}_{i}$) from a non-symmetric Lanczos algorithm using the $1^{+}$
subspace of $\overline{H}_N$ and rewrite Eq.~(\ref{M2vbb_sum}) as a
continued fraction~\cite{engel1992}. This computation typically
requires about 10-20 Lanczos iterations. With the
similarity-transformed operator,
$\overline{O}=\overline{\mathbf{\sigma}\tau^{-}}$, and the pivot
states $\langle\nu_{\mathrm{F}}\vert=\langle\Phi_0 \vert
L\overline{O}$,
$\vert\nu_{\mathrm{I}}\rangle=\overline{O}\vert\Phi_0\rangle$,
$\langle\nu_{\mathrm{I}}\vert=\langle\Phi_0 \vert ( 1+\hat{\Lambda}
)\overline{O^{\dagger}}$, and
$\vert\nu_{\mathrm{F}}\rangle=\overline{O^{\dagger}}
R\vert\Phi_0\rangle$, the NME becomes
\begin{equation}
  \label{M2vbb_lanczos}
  \medmuskip=3mu
  \vert M^{2\nu} \vert^{2} = \frac{\langle\nu_{\mathrm{F}}\vert\nu_{\mathrm{I}}\rangle}{a_{0}+\frac{E_{\mathrm{I}}-E_{\mathrm{F}}}{2}-\frac{b_{0}^{2}}{a_{1}+\cdots}}
  \frac{\langle\nu_{\mathrm{I}}\vert\nu_{\mathrm{F}}\rangle}{a^{*}_{0}+\frac{E_{\mathrm{I}}-E_{\mathrm{F}}}{2}-\frac{(b^{*}_{0})^{2}}{a^{*}_{1}+\cdots}}.
\end{equation}

\begin{figure}
  \includegraphics[width=0.5\textwidth]{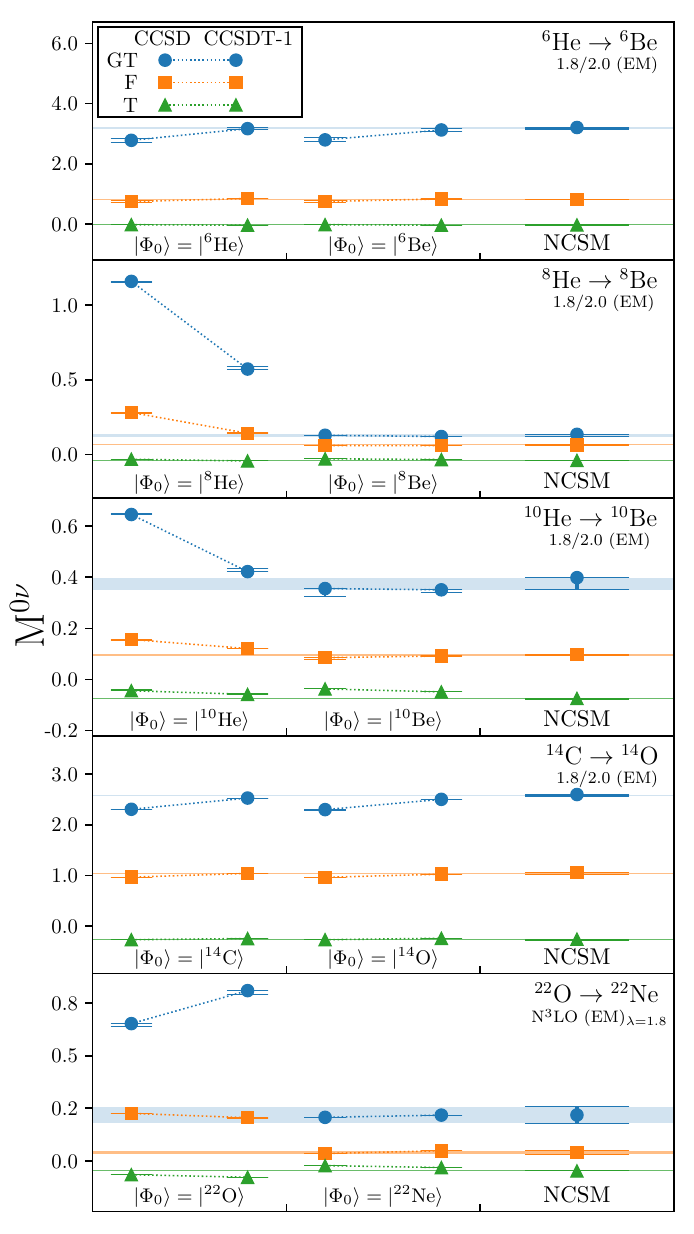}
  \caption{(Color online) Comparison of the $0\nu\beta\beta$ NME in several light
    nuclei computed with the coupled cluster method and the no-core shell model.
    The first two columns correspond to different choices for the coupled-cluster
    reference state, and results from the CCSD and CCSDT-1 approximations are shown
    in each. The error bars indicate the uncertainties coming from variations with
    model-space size. Each case utilizes the 1.8/2.0 (EM) interaction
    except for $^{22}$O$\to^{22}$Ne which disregards the three-nucleon forces to
    more rapidly converge the NCSM results.} \label{benchmark-0vbb}
\end{figure}

{\it Benchmarks.--- } To gauge the quality of our coupled-cluster
computations we benchmark with the more exact no-core shell model
(NCSM)~\cite{navratil2000,navratil2009,barrett2013} by computing
$0\nu\beta\beta$ matrix elements in light nuclei.  Although the
$0\nu\beta\beta$ decay of these isotopes are energetically forbidden
or would be swamped by successive single-$\beta$ decays in an
experiment, the benchmarks still have theoretical value.
Figure~\ref{benchmark-0vbb} shows the $0\nu\beta\beta$ matrix elements
of the GT, F, and T operators for the transitions $^{6}$He$\to^{6}$Be,
$^{8}$He$\to^{8}$Be, $^{10}$He$\to^{10}$Be, $^{14}$C$\to^{14}$O, and
$^{22}$O$\to^{22}$Ne. The coupled-cluster results are shown in pairs,
with both the initial and final state as the reference.  For each
pair, the first (second) point shows the CCSD (CCSDT-1) approximation;
these two points are connected by dotted lines. The vertical error
bars indicate the change of the matrix element as the model space is
increased from $N_{\rm max}=8$ to $N_{\rm max}=10$.  The NCSM results
are shown in the third column, and their error bars indicate
uncertainties from extrapolation to infinite model spaces. The shaded
bands are simply to facilitate comparison.

The NMEs in the mirror-symmetric cases $^{6}$He$\to^{6}$Be and
$^{14}$C$\to^{14}$O depend very little (within about $1\%$) on the
choice of the initial or final nucleus as the reference state, a
result that is consistent with the weak charge-symmetry breaking of
the chiral interaction. For the $A=14$ transition between doubly
closed-shell nuclei, coupled-cluster theory and NCSM results agree
within about $3\%$.  The small contributions of triples correlations
($< 10\%$) suggest that these results are accurate. The results are of
similar quality for $^{6}$He$\to^{6}$Be, even though these nuclei are
only semi-magic. The case of $^{10}$He$\to^{10}$Be is slightly more
challenging, with a doubly closed-shell initial nucleus and a
partially closed-shell final nucleus. Comparing our results for
$^{6}$He$\to^{6}$Be with other works is complicated by the
lack of renormalization-group invariance. However,
\textcite{pastore2018,cirigliano2019} found absolute values that are
similar to ours using a harder interaction, and \textcite{basili2020}
also agrees with our results (apart from an arbitrary sign), although
they did not include three-nucleon forces.

The cases of $^{8}$He$\to^{8}$Be and $^{22}$O$\to^{22}$Ne are more
challenging still, because the final nuclei are truly open-shell
systems.  Adding triples correlations to the spherical results induces
a $\sim 50\%$ change in the first case and worsens the agreement with
NCSM in the second, suggesting the need for more particle-hole
excitations. Once again, however, using the deformed final state as
the reference leads to results that are both consistent with the NCSM
and converged at the CCSDT-1 level. Thus, the coupled-cluster results
are more accurate when the open-shell (or deformed) nucleus is taken
as the reference, and they agree within smaller model-space
uncertainties with the NCSM benchmarks.

The benchmark calculations suggest that the two approaches (with a
spherical $^{48}$Ca or a deformed $^{48}$Ti as the reference state)
allow us to bracket the NME. The result from the first approach
exceeds the exact NME because the imposition of spherical symmetry
increases the overlap of the initial and final wave functions. The
second result underestimates the exact NME, probably because the
deformations of the initial and final states are quite
different. Generator-coordinate
methods~\cite{rodriguez2011} might have an advantage here, and we
expect that symmetry projection would make the results more
accurate.

Unfortunately, we are not able to extend the benchmarks to
heavier nuclei. Benchmarks with the traditional shell model are
complicated because coupled-cluster theory in its singles, doubles,
and triples approximation does not accurately capture the strong
correlations in small shell-model spaces~\cite{horoi2007}, see
Supplemental Material for more details.

Although the coupling strength of the leading-order contact potential
in the $0\nu\beta\beta$ operator is
unknown~\cite{cirigliano2018,cirigliano2019,cirigliano2020}, we
attempt to estimate its effect by applying the
coupled-cluster methods discussed above with the addition of a contact
term, $V_c(\mathbf{r_{12}})=2\pi^2 g\delta(\mathbf{r}_{12})\tau_-^{(1)}\tau_-^{(2)}$,
to the operator, $\hat{O}_{0\nu}$. Using a coupling strength of
$g=\pm 1$~fm$^2$ results in a
NME of $0.15 \le M^{0\nu} \le 1.02$ (see Supplemental Material for
details).

{\it Two-neutrino double-beta decay of $^{48}$Ca.--- } The
$2\nu\beta\beta$ decay of $^{48}$Ca was accurately predicted by
\textcite{caurier1990} before its
observation~\cite{balysh1996,brudanin2000,arnold2016}.  Subsequent
authors studied this decay
further~\cite{horoi2007b,raduta2011,horoi2013}, and evaluations can be
found in Refs.~\cite{barabash2015,barabash2019}. We compute the matrix
element for the $2\nu\beta\beta$ decay of $^{48}$Ca with the 1.8/2.0
(EM) interaction and the Lanczos continued fraction method. We employ
a spherical $^{48}$Ca natural-orbital basis and converge our results
with respect to $N_{\rm max}$ and the number of $3p$--$3h$
configurations included in the wave functions of $^{48}$Ca, $^{48}$Ti,
and the intermediate nucleus $^{48}$Sc. The results are also converged
with respect to the number of Lanczos iterations used in the continued
fraction~(\ref{M2vbb_lanczos}). We note that the $2\nu\beta\beta$
calculations can only be performed in the spherical scheme since we
sum over intermediate states with definite spin.

\begin{figure}
  \includegraphics[width=0.5\textwidth]{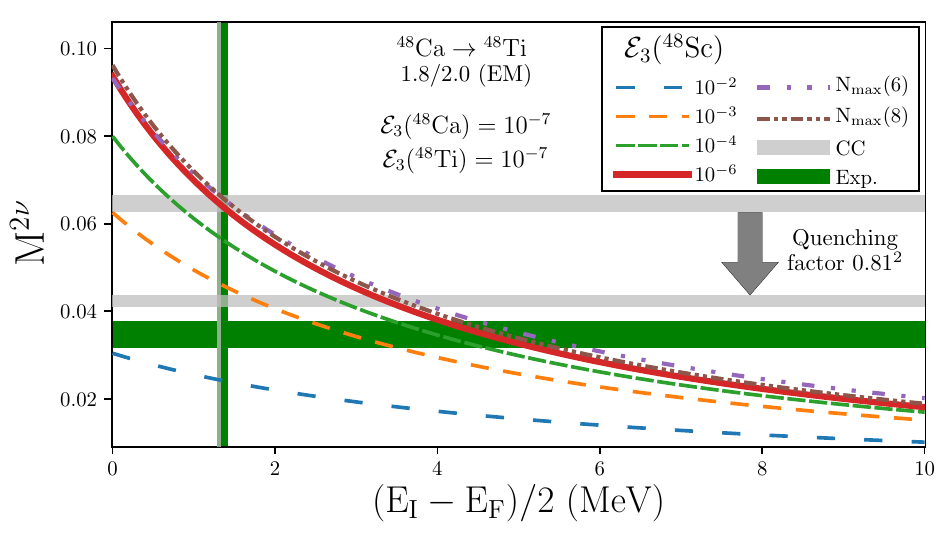}
  \caption{(Color online) The NME for the $2\nu\beta\beta$ decay
    $^{48}$Ca$\to^{48}$Ti computed with the 1.8/2.0 (EM) interaction
    as a function of the energy difference, $E_{\mathrm{I}}-E_{\mathrm{F}}$,
    and the $3p$--$3h$ truncation used to calculate $^{48}$Sc,
    $\mathcal{E}_{3}$, at $N_{\rm max} = 10$. The results for
    $N_{\rm max} = 6,8$ are also shown. The experimental NME and
    energy difference are are shown along with the computed
    energy difference and NME, with and without a quenching factor
    of $0.81^2$ deduced from two-body currents~\cite{gysbers2019}.}
  \label{2vbb-48ca}
\end{figure}

Figure~\ref{2vbb-48ca} shows the NME for the $2\nu\beta\beta$ decay of
$^{48}$Ca, computed in the CCSDT-1 approximation, as a function the energy
difference, $E_{\mathrm{I}}-E_{\mathrm{F}}$, with different curves representing
both the $N_{\rm max}$ convergence and $\mathcal{E}_{3}$ convergence of
$^{48}$Sc. The converged result, $M^{2\nu}=0.065 \pm 0.002$, is at the
intersection with the theoretical energy difference between the ground-state
energies of $^{48}$Ca and $^{48}$Ti computed from the corresponding reference
states, $(E_{\mathrm{I}}-E_{\mathrm{F}})/2=1.32$~MeV. Given that $E$ is equivalent
to the negative binding energy, $E=-BE$, this is consistent with
the experimental difference, $[BE(^{48}\mathrm{Ti})-BE(^{48}\mathrm{Ca})]/2=1.35$~MeV.
The uncertainty in our result represents the error from the different convergence
criteria. These results are sensitive to the energy of the first $1^{+}$
state in $^{48}$Sc. Our value of $\Delta E_{\mu=0}=2.93$~MeV is close to
the corresponding experimental value of
$BE(^{48}\mathrm{Ca})-BE(^{48}\mathrm{Sc}_{\mu=0}^{1^{+}})=3.02$~MeV,
and the NME gets reduced by about 2\% if one uses the experimental
datum instead. The comparison of the values in Eq.~(\ref{M2vbb_sum})
to experiment are detailed in the Supplemental Material.

We multiply our matrix element with the a quenching factor $q^{2} =
0.81^{2}$ deduced from two-body currents in a recent coupled-cluster
computation of the Ikeda sum-rule in $^{48}$Ca~\cite{gysbers2019} which
includes all final $1^+$ states in $^{48}$Sc and is similar to Eq.~(\ref{M2vbb_sum}). We
obtain $q^2 M^{2\nu}=0.042 \pm 0.001$ which is somewhat larger than
the experimental value of $M^{2\nu}= 0.035 \pm 0.003$
~\cite{barabash2019,stoica2013}. This is most likely due to our
inability to accurately describe the deformed nature of $^{48}$Ti.
In a future work we will
investigate the role of momentum dependent two-body currents on
this decay. We note that the quenching factor from the Ikeda sum-rule
weights all $1^+$ states equally (as there is no energy denominator) and
is somewhat larger than the phenomenological value of $q^{2} = 0.74^{2}$~\cite{martinezpinedo1996}.
We verified our methods by performing two $2\nu\beta\beta$ benchmarks,
of $^{48}$Ca in the $pf$-shell and of $^{14}$C in a full no-core model
space, which are shown in the Supplemental Material. The former is
compared with exact diagonalization, and the latter with the NCSM.

{\it Conclusions.---} Using interactions from chiral EFT and the
coupled-cluster method, we computed the nuclear matrix elements for
0$\nu\beta\beta$-decay of $^{48}$Ca$\to^{48}$Ti and found a relatively
small value. The uncertainties stem from the treatment of nuclear
deformation and are supported by extensive benchmarks.
We also calculated the 2$\nu\beta\beta$-decay of
$^{48}$Ca$\to^{48}$Ti and included the ab-initio quenching factor from
two-body currents of the Ikeda sum-rule in $^{48}$Ca.

\begin{acknowledgments}
  We thank A. Belley, V. Cirigliano, J. de Vries, H. Hergert,
  J. D. Holt, M. Horoi, J. Men{\'e}ndez, C. G. Payne, S. R. Stroberg,
  A. Walker-Loud, and J. M. Yao, for useful discussions.  This work
  was supported by the Office of Nuclear Physics, U.S.  Department of
  Energy, under Grants DE-FG02-96ER40963, DE-FG02-97ER41019
  DE-SC0008499 (NUCLEI SciDAC collaboration), the Field Work Proposal
  ERKBP57 at Oak Ridge National Laboratory (ORNL) and SCW1579 at
  Lawrence Livermore National Laboratory (LLNL), the National Research
  Council of Canada, and NSERC, under Grants SAPIN-2016-00033 and
  PGSD3-535536-2019. TRIUMF receives federal funding via a
  contribution agreement with the National Research Council of Canada.
  This work was prepared in part by LLNL under Contract
  No.~DE-AC52-07NA27344. Computer time was provided by the Innovative
  and Novel Computational Impact on Theory and Experiment (INCITE)
  program. This research used resources of the Oak Ridge Leadership
  Computing Facility located at ORNL, which is supported by the Office
  of Science of the Department of Energy under Contract
  No.~DE-AC05-00OR22725.
\end{acknowledgments}


%

\clearpage
\newpage

\appendix*

\section{Supplemental Material: Coupled-Cluster Calculations of Neutrinoless Double-$\beta$ Decay in $^{48}$Ca}

\subsection{Composition of $0\nu\beta\beta$ matrix elements}
\begin{figure}[h]
  \includegraphics[width=0.5\textwidth]{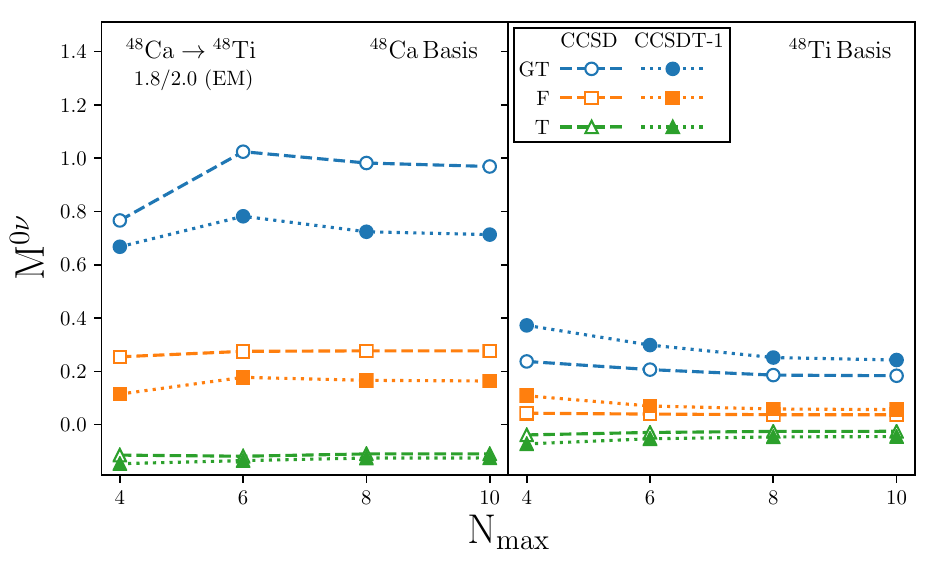}
  \caption{Different components of the NME for the
    $0\nu\beta\beta$ decay of $^{48}$Ca using both the CCSD and CCSDT-1
    approximations with both spherical and deformed reference states.
    The results are converged with respect to $N_{\rm max}$.}
  \vspace{-0.75em}
  \label{ca48-0vbb-components}
\end{figure}

The nuclear matrix elements of the the $0\nu\beta\beta$ decay
$^{48}$Ca$\to^{48}$Ti consists of three contributions [the
  Gamow-Teller (GT), Fermi (F), and tensor (T) terms], and we have
$M_{0\nu} =M^{\mathrm{GT}}_{0\nu} + M^{\mathrm{F}}_{0\nu} +
M^{\mathrm{T}}_{0\nu}$. The individual contributions are shown in
Fig.~\ref{ca48-0vbb-components} in the CCSD and CCSDT-1 approximations
computed in spherical and deformed natural-orbital bases. The
potential is the 1.8/2.0 (EM) interaction~\cite{hebeler2011}, and
results are shown as a function of the model-space size $N_{\rm max}$.
The results for $N_{\rm max}=10$ correspond to the full NMEs shown in
Fig.~\ref{ca48-0vbb}. Similarly to \textcite{belley2020} we find a
sizeable tensor component.

\subsection{Benchmarks for energies of light nuclei}
\begin{figure}[h]
  \includegraphics[width=0.5\textwidth]{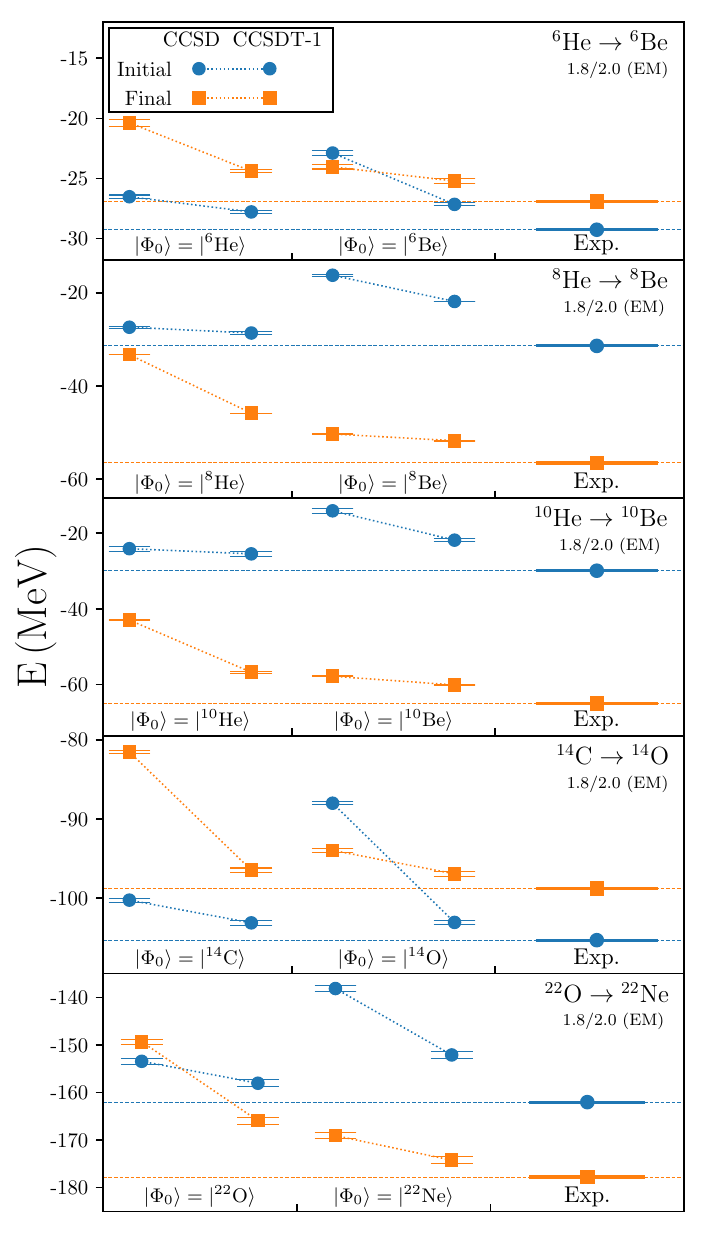}
  \caption{Comparison of the ground-state energies
    for the several light nuclei involved in our $0\nu\beta\beta$
    benchmark calculations with their experimental values. The
    first two columns indicate which nucleus was taken as the
    reference state, and results from the CCSD and CCSDT-1
    approximations are shown. The error bars indicate the
    uncertainties with respect to the model-space size.}
  \vspace{-0.75em}
  \label{benchmark-energy}
\end{figure}

We also computed the ground-state energies for the benchmark nuclei
$^{6}$He$\to^{6}$Be, $^{8}$He$\to^{8}$Be, $^{10}$He$\to^{10}$Be,
$^{14}$C$\to^{14}$O, and $^{22}$O$\to^{22}$Ne. Figure~\ref{benchmark-energy}
shows the results from coupled-cluster CCSD and CCSDT-1 computations
and compares them to data for the 1.8/2.0 (EM) interaction. We remind
the reader that this interaction yields accurate binding energies across
the lower half of the nuclear chart. As indicated, the coupled-cluster
results used both the initial and final nuclei as reference states.
While deformed reference states were sufficient to match the NCSM results
for the $0\nu\beta\beta$ nuclear matrix elements shown in
Fig.~\ref{benchmark-0vbb} of the main text, the ground-state energies
are underbound by a few MeV which are expected to be obtained when
restoring the broken spherical symmetry.

\subsection{Spectrum of $^{48}$Ti}
\begin{figure}[h]
  \includegraphics[width=0.5\textwidth,trim=5 0 0 0,clip]{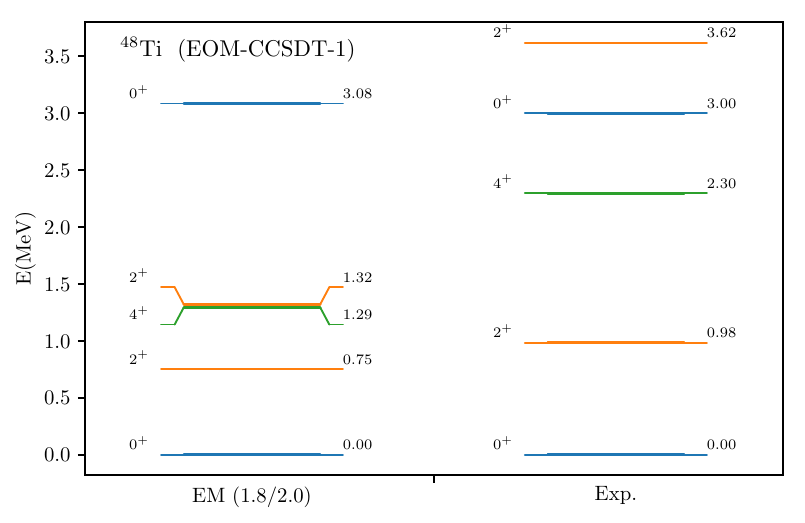}
  \caption{Energies of low-lying states in $^{48}$Ti
    with respect to the ground states using the 1.8/2.0 (EM) interaction
    compared with experiment. These results use the EOM-CCSDT-3 approximation
    with a spherical $^{48}$Ca Hartree-Fock reference state (see text for details).}
  \vspace{-0.75em}
  \label{titanium-spectra}
\end{figure}

Because of the strong correlation between the accuracy of the
$0\nu\beta\beta$ NME and the quality of the excitation spectra of the
initial and final nuclei, we calculate the excitation spectrum of
$^{48}$Ti with the double-charge exchange EOM-CCSDT-3 approximation
using a spherical $^{48}$Ca Hartree-Fock basis. The spectrum for the
1.8/2.0 (EM) interaction is shown in Fig.~\ref{titanium-spectra} and
compared with experiment.

The compressed $2^{+}$ and $4^{+}$ states of the 1.8/2.0 (EM) spectrum
show that the triples correlations in a spherical basis are
insufficient to represent the deformed nucleus and motivates the usage
of deformed reference states.

\subsection{Additional $2\nu\beta\beta$ decay material}
\begin{figure}[h]
  \includegraphics[width=0.5\textwidth]{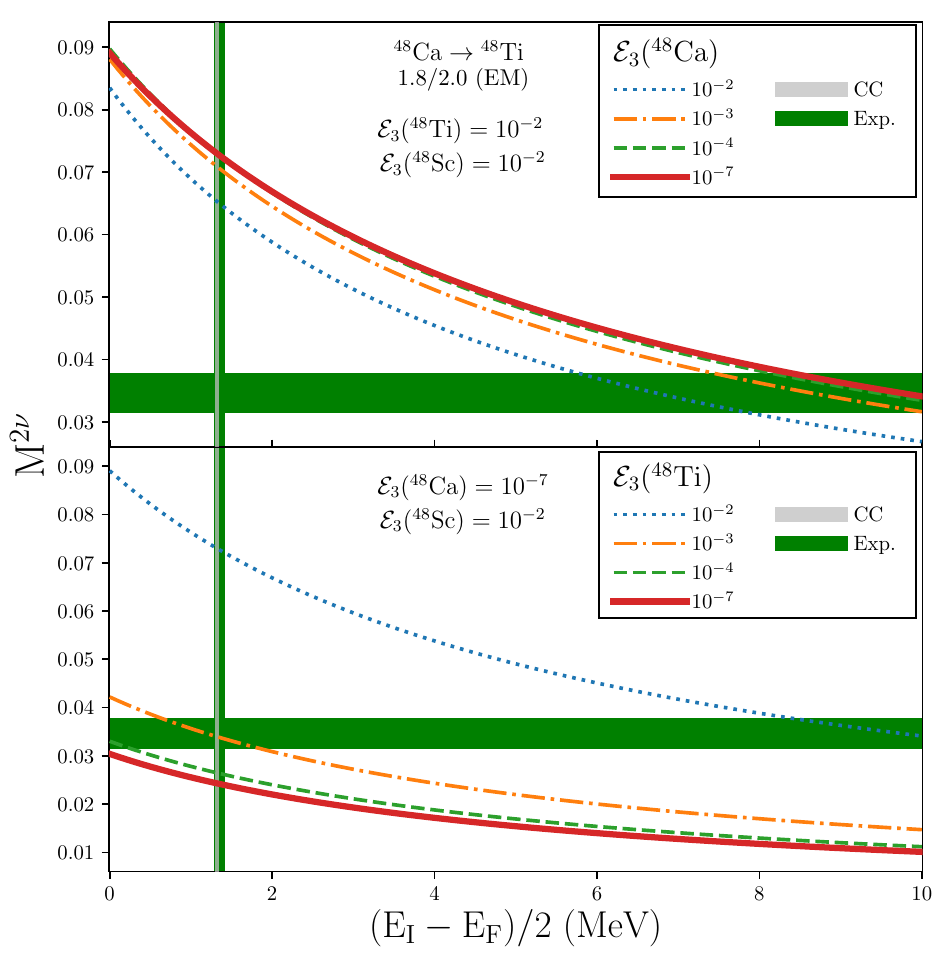}
  \caption{The NME for the $2\nu\beta\beta$ decay
    $^{48}$Ca$\to^{48}$Ti computed with the Lanczos method and
    the 1.8/2.0 (EM) interaction as a function of the energy
    difference, $E_{\mathrm{I}}-E_{\mathrm{F}}$, and the
    $3p$--$3h$ truncation, $\mathcal{E}_{3}$, used to calculate
    $^{48}$Ca (top) and $^{48}$Ti (bottom). The results use the
    CCSDT-1 approximation and $N_{\rm max} = 10$. The experimental
    NME is shown in horizontal bands while the experimental and
    computed energy difference are are shown in vertical bands.}
  \vspace{-0.75em}
  \label{2vbb-48ca-supp}
\end{figure}

The convergence of the NME for the $2\nu\beta\beta$ decay of $^{48}$Ca
with respect to the the $3p$--$3h$ truncation, $\mathcal{E}_{3}$, is computed
for the initial nucleus, $^{48}$Ca, the final nucleus, $^{48}$Ti, and the
intermediate nucleus, $^{48}$Sc, successively. The latter is shown in
Figure~\ref{2vbb-48ca}, and the former two are shown in Figure~\ref{2vbb-48ca-supp}.
These calculations utilize the CCSDT-1 approximation in a spherical $^{48}$Ca natural
orbital basis with the 1.8/2.0 (EM) interaction. Not shown is the convergence with
respect to the number of iterations used in the Lanczos (continued fraction) method.
Our final results need only 20 Lanczos iterations which converges very rapidly and
does not contribute to the uncertainty.

The denominator in Eq.~(\ref{M2vbb_sum}) and~\cite{vogel2012,kotila2012} can
be defined conceptually as the total energy difference between the excited-state
intermediate nucleus and the average of the initial and final nuclei, including
electron masses, $D=M_{\mu}-(M_{\mathrm{I}}+M_{\mathrm{F}})/2$, where $M$ is the
atomic mass. However, this can be written in several ways depending on the context.
First, because the EOM-CC method that we employ gives the intermediate energies with
respect to the initial nucleus, it's natural for us to rewrite this accordingly,
$D=M_{\mu}-M_{\mathrm{I}}+(M_{\mathrm{I}}-M_{\mathrm{F}})/2$, where
$M_{\mathrm{I}}-M_{\mathrm{F}}$ is the double beta decay $Q$ value, $Q_{\beta\beta}$.
Next, because we neglect the neutron-proton mass difference and electron masses in
our calculations, we can cancel these contributions from each term in the denominator,
$D=E_{\mu}-E_{\mathrm{I}}+(E_{\mathrm{I}}-E_{\mathrm{F}})/2$, where $E$ is equivalent
to the negative binding energy, $E=-BE$. This is the form of the denominator that
we employ in Eq.~(\ref{M2vbb_sum}), $D=\Delta E_{\mu}+(E_{\mathrm{I}}-E_{\mathrm{F}})/2$.
By adding the appropriate nucleon and electron masses, we can compare our results
to direct experimental values for the $2\nu\beta\beta$ decay of $^{48}$Ca; for the
first $1^{+}$ state in $^{48}$Sc, $M_{\mu=0}-M_{\mathrm{I}}=1.73$~MeV compared to our
value of 1.64~MeV, and the experimental value for $Q_{\beta\beta}=4.27$~MeV is compared
to our value of 4.20~MeV.

\begin{figure}[h]
  \includegraphics[width=0.5\textwidth]{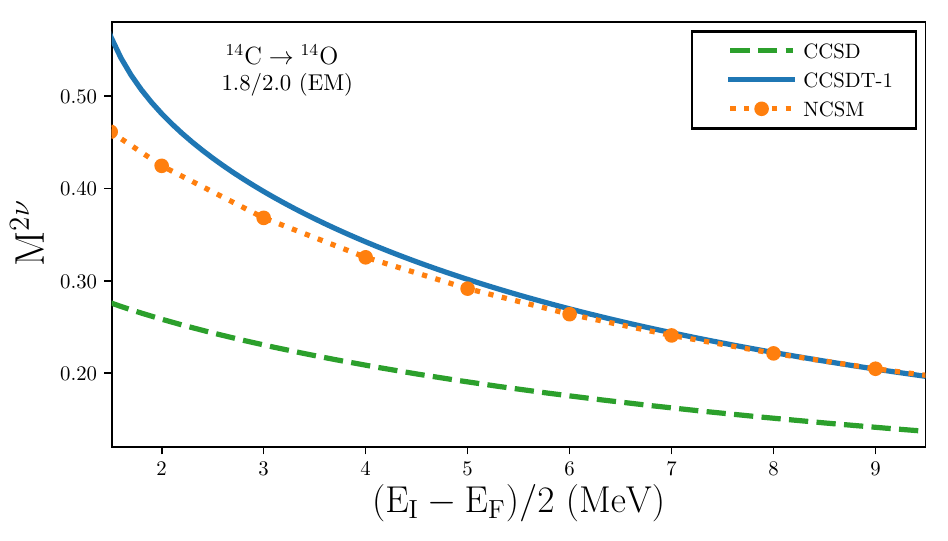}
  \caption{Comparison of the NME for the
    $2\nu\beta\beta$ decay of $^{14}$C$\to^{14}$O computed with the
    no-core shell model and coupled cluster at both the CCSD and
    CCSDT-1 approximations. All results use the Lanczos continued
    fraction method, and the CCSDT-1 results include all $3p$--$3h$
    configurations and are converged at $N_{\rm max} = 10$.}
  \vspace{-0.75em}
  \label{C14-benchmark-2vbb}
\end{figure}

We perform an additional benchmark for the fictitious $2\nu\beta\beta$
decay of $^{14}$C$\to^{14}$O by comparing our results to the no-core
shell model in a full model space using the 1.8/2.0 (EM)
interaction. Both methods use the Lanczos continued fraction method
and are converged with respect to $N_{\rm max}$. Given the relatively
small size of the calculations, the CCSDT-1 results include all
$3p$--$3h$ configurations. Additionally, the coupled cluster results
are computed in a spherical $^{14}$C natural orbital basis. These
results, shown in Figure~\ref{C14-benchmark-2vbb}, once again bolster
the validity of the Lanczos method applied within coupled cluster
theory, and shows the importance of including $3p$--$3h$
configurations in these calculations.

\begin{figure}[h]
  \includegraphics[width=0.5\textwidth]{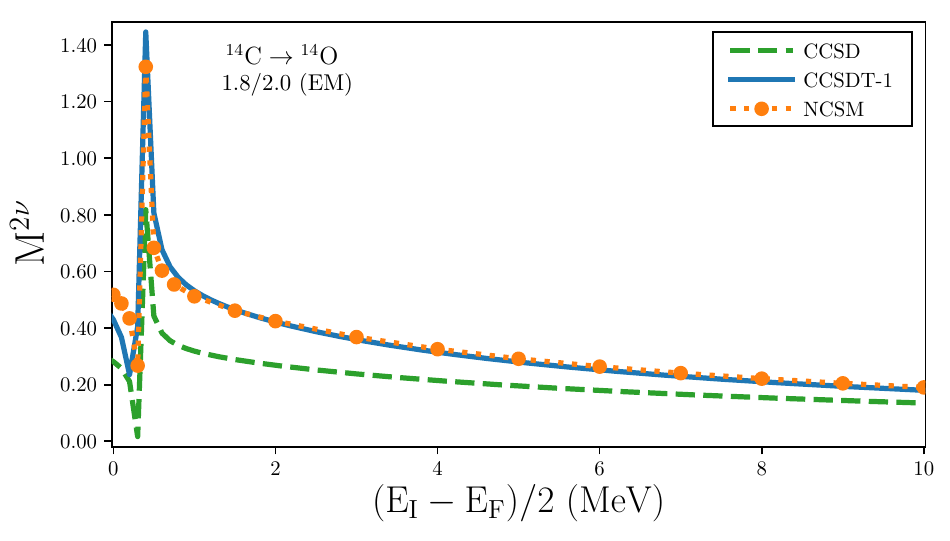}
  \caption{Comparison of the NME for the $2\nu\beta\beta$ decay of
    $^{14}$C$\to^{14}$O computed with the no-core shell model and coupled cluster
    at both the CCSD and CCSDT-1 approximations. All results use the Lanczos
    continued fraction method, and the CCSDT-1 results include all $3p$--$3h$
    configurations and are converged at $N_{\rm max} = 10$. Each curve is
    shifted so that the singularity corresponds to the experimental value
    for $E(^{14}\mathrm{C})-E(^{14}\mathrm{N})$.}
  \vspace{-0.75em}
  \label{C14-benchmark-2vbb2}
\end{figure}

The shapes of the curves in Figure~\ref{C14-benchmark-2vbb} capture
the spectra of $1^{+}$ states relative to the $1^{+}$ ground state in
$^{14}$N. The absolute position of these curves is determined by the
difference in ground state energies between the initial and intermediate
nucleus, which corresponds to the first pole in the Green's function
$1/(z+\overline{H}_N)$ and is marked by the first singularity on the
curve. To properly compare NMEs taken from different curves we shift
each singularity to the experimental value, which is shown in
Figure~\ref{C14-benchmark-2vbb2}.

\subsection{Benchmarks with the traditional shell model}
We also performed computations for $pf$ shell nuclei and compared with
exact results from shell-model calculations in the $pf$
shell~\cite{menendez2014} based on the interactions
GXPF1A~\cite{honma2004} and KB3G~\cite{poves2001}. We caution,
however, that coupled-cluster theory with singles, doubles, and
triples excitations might not be accurate in traditional shell-model
spaces. \textcite{horoi2007} found that a small shell gap makes it
necessary to include many-particle--many-hole correlations.

\begin{figure}[h]
  \includegraphics[width=0.5\textwidth]{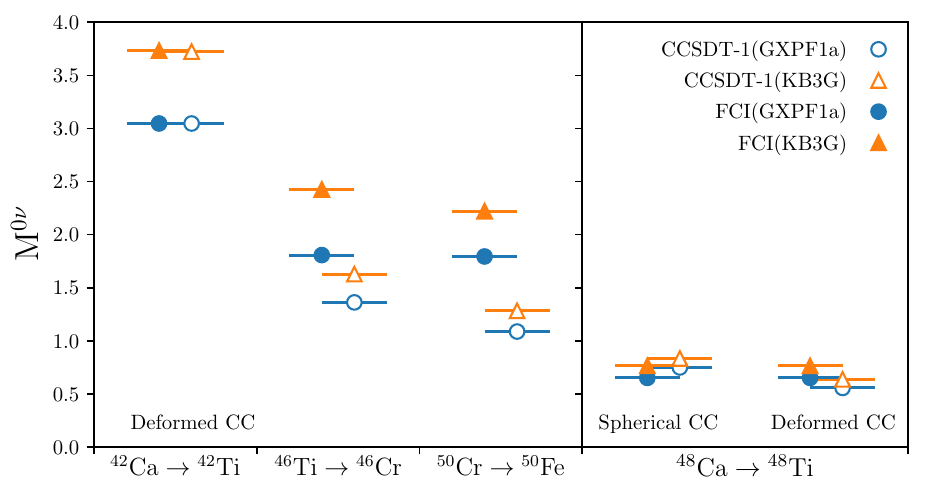}
  \caption{Comparison in several $pf$-shell of the
    $0\nu\beta\beta$ NMEs between CCSDT-1 and exact shell-model
    calculations, with the GXPF1A and KB3G interactions.  In
    $^{42}$Ca, $^{46}$Ti, and $^{50}$Cr we use a deformed reference
    state in the initial nucleus, while for the decay $^{48}$Ca$\to
    ^{48}$Ti we use reference states in both nuclei.}
  \vspace{-0.75em}
  \label{0vbb-pfshell}
\end{figure}

Figure~\ref{0vbb-pfshell} shows CCSDT-1 results and compares them to
exact results for the transitions $^{42}$Ca$\to^{42}$Ti,
$^{46}$Ti$\to^{46}$Cr, $^{50}$Cr$\to^{50}$Fe, and
$^{48}$Ca$\to^{48}$Ti.  In the first of these, the valence shell
contains only two nucleons and the problem is thus exactly solvable
with CCSD.  The next two cases are particularly challenging because
initial and final nuclei are open-shell systems.  Here,
coupled-cluster NMEs are significantly smaller than their exact
counterparts.  For the most relevant case, $^{48}$Ca$\to^{48}$Ti,
coupled-cluster results are $\sim 15\%$ lower (higher) than the
benchmarks when $^{48}$Ti ($^{48}$Ca) serves as the reference,
i.e.\ when we use a deformed (spherical) reference state.

In Fig.~\ref{0vbb-pfshell}, the results for $^{48}$Ca$\to ^{48}$Ti
seem again to suggest that the spherical and deformed coupled-cluster
calculations are bracketing the exact benchmarks. This, however, is
not true in general, as calculations for the $0\nu\beta\beta$ NMEs in
$^{52,54}$Ca show.

\begin{figure}[h]
  \includegraphics[width=0.5\textwidth]{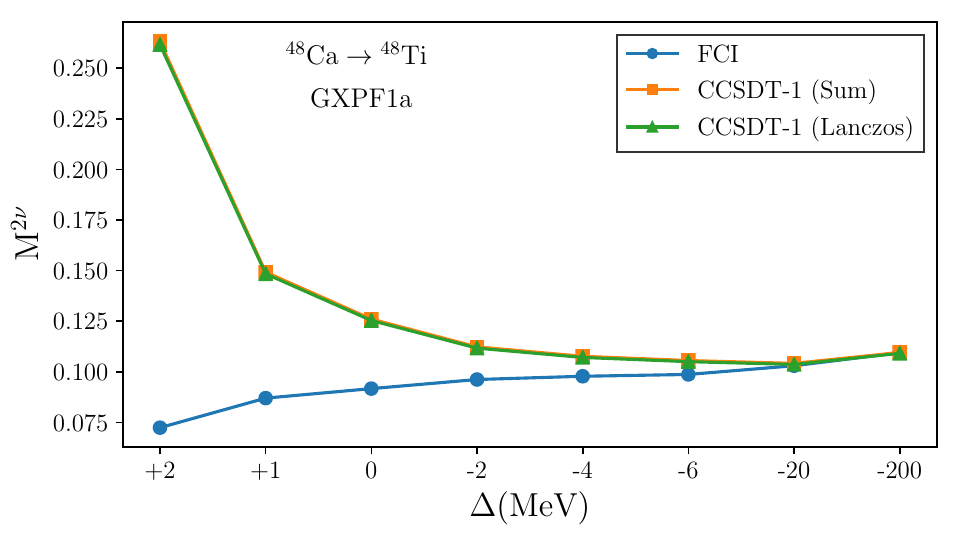}
  \caption{Comparison of the nuclear matrix element for
    the $2\nu\beta\beta$ decay of $^{48}$Ca$\to^{48}$Ti between
    CCSDT-1 and exact shell-model calculation, computed in the
    $pf$-shell with the GXPF1A interaction as a function of the gap
    between the $f_{7/2}$ and $p_{3/2}$ shells, where $\Delta=0$
    corresponds to the phenomenological value of the GXPF1A
    interaction. The CC results are shown using both the explicit
    summation from Eq.~(\ref{M2vbb_sum}) and the Lanczos method of
    Eq.~(\ref{M2vbb_lanczos}).}
  \vspace{-0.75em}
  \label{pfshell2vbb}
\end{figure}

To benchmark our $2\nu\beta\beta$ decay results of $^{48}$Ca, we
compare NMEs computed with coupled-cluster in the CCSDT-1
approximation with exact results from shell-model calculations in the
$pf$ shell phenomenological interaction GXPF1A~\cite{honma2004}.
Figure~\ref{pfshell2vbb} shows the NME as a function of the the energy
gap between the $f_{7/2}$ and $p_{3/2}$ shells, $\Delta$. The original
GXPF1A interaction is given by $\Delta=0$, and $\Delta\rightarrow
-\infty$ minimizes any correlations, which essentially makes the exact
shell-model method equivalent to the approximate coupled-cluster
method~\cite{horoi2007}. To properly compare the different methods,
$E_{\mathrm{I}}-E_{\mathrm{F}}$ and $\Delta E_{\mu=0}$ in the
denominator of Eq.~(\ref{M2vbb_sum}) are fixed to their experimental values.

In Fig.~\ref{pfshell2vbb} we also compare the Lanczos method to the
explicit sum over intermediate $1^{+}$ states in $^{48}$Sc as in
Eq.~(\ref{M2vbb_sum}). For these results, the Lanczos method used only
20 iterations while the summation used 60 intermediate states which
required $\sim 300$ iterations. These results confirm the validity of
the Lanczos method and the validity of the coupled-cluster method for
the $2\nu\beta\beta$ NME when important correlations are included.

\begin{figure}[h]
  \includegraphics[width=0.5\textwidth,trim=20 0 0 0,clip]{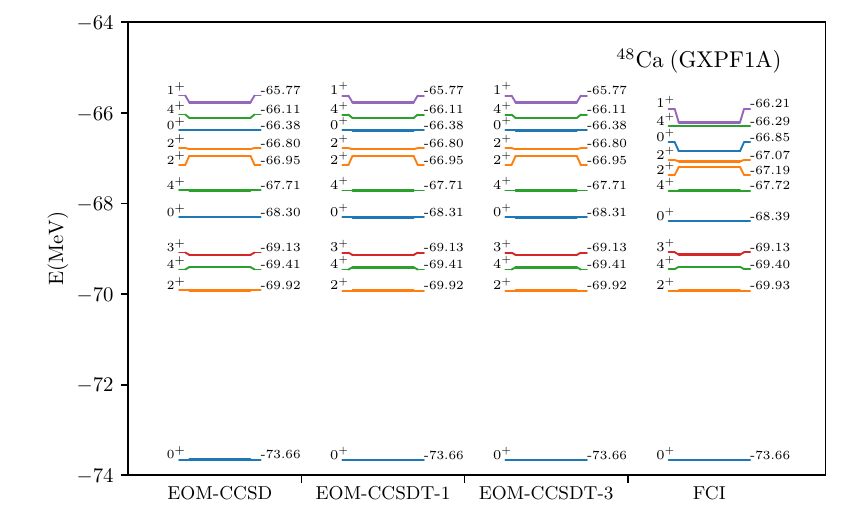}
  \caption{Energies of low-lying states in $^{48}$Ca
    with respect to the ground states using the GXPF1A interactions
    in the pf-shell compared with full diagonalization (FCI). These
    results use the EOM-CCSD, EOM-CCSDT-1, and EOM-CCSDT-3 approximations with
    a spherical $^{48}$Ca Hartree-Fock reference state (see text for details).}
  \vspace{-0.75em}
  \label{calcium-spectra-gxpf}
\end{figure}

\begin{figure}[h]
  \includegraphics[width=0.5\textwidth,trim=20 0 0 0,clip]{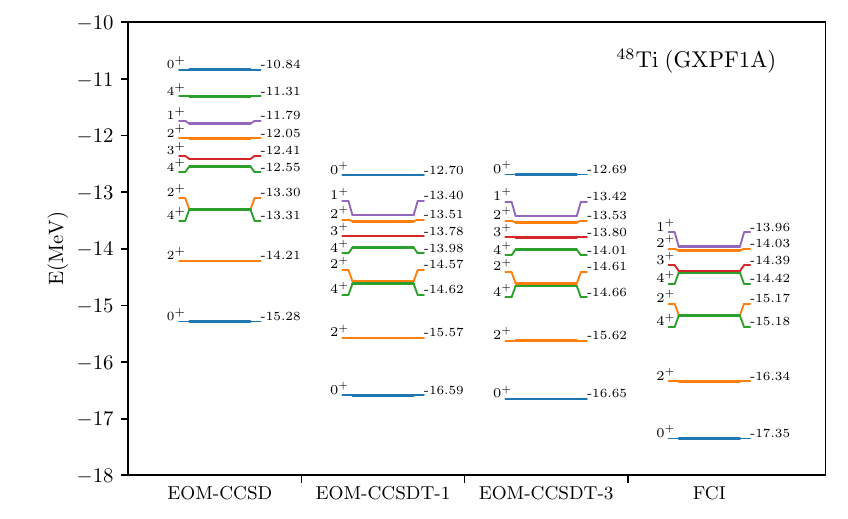}
  \caption{Energies of low-lying states in $^{48}$Ti
    with respect to the ground states using the GXPF1A interactions in
    the pf-shell compared with full diagonalization (FCI). These
    results use the EOM-CCSD, EOM-CCSDT-1, and EOM-CCSDT-3
    approximations with a spherical $^{48}$Ca Hartree-Fock reference
    state (see text for details).}
  \vspace{-0.75em}
  \label{titanium-spectra-gxpf}
\end{figure}

We also calculated the low-lying spectrum in $^{48}$Ca and $^{48}$Ti
with a spherical $^{48}$Ca Hartree-Fock basis. The results are shown
in Figs.~\ref{calcium-spectra-gxpf} and \ref{titanium-spectra-gxpf},
respectively, using the EOM-CCSD, EOM-CCSDT-1, and the EOM-CCSDT-3
approximations~\cite{urban1985,noga1987,shavittbartlett2009}.
The triples approximations do not add any binding energy for $^{48}$Ca
because there are no $3p$--$3h$ configurations for this nucleus in the
$pf$-shell.

With triples contributions included, the spectra of both, $^{48}$Ca
and $^{48}$Ti, agree with the exact diagonalization. The closed-shell
nucleus $^{48}$Ca is well-described already in the EOM-CCSD
approximation. As the restricted model space does not allow for any
$3p$--$3h$ configurations, the spectrum does not change in the
EOM-CCSDT-1 or EOM-CCSDT-3 approximations. The nucleus $^{48}$Ti is
computed with the double-charge exchange EOM-CC. Despite the quality
of both these spectra, the ground-state energy of $^{48}$Ti is less
accurate than that of $^{48}$Ca, and the nuclear matrix element (shown
in Fig.~\ref{0vbb-pfshell}) deviates by about $\sim 15\%$ from the
exact result. This reflects the sensitivity of the this matrix element
with respect to the spectra of the initial and final
nuclei~\cite{jiao2018}.

\subsection{Note on the $0\nu\beta\beta$ contact}
For the contact we use a separable potential from a discrete variable
representation
(DVR)~\cite{light1985,baye1986,light2007,littlejohn2002,bulgac2013} in
the harmonic oscillator (HO) basis with matrix elements
\be \langle n' 0|V|n 0\rangle = \sum_{\mu=0}^N c_{\mu,0}^2\tilde{\psi}_{n',
  0}(k_{\mu, 0}) \sum_{\nu=0}^N c_{\nu,0}^2\tilde{\psi}_{n, 0}(k_{\nu, 0})
\ .
\label{HOME}
\ee
Here, HO eigenstates with radial quantum number $n$ and angular
momentum $l$ are denoted as as $|n l\rangle$.  The basis is finite,
i.e. $n, n' =0, 1, \ldots, N$. We used $\tilde{\psi}_{\mu, l}(k)$ for
the radial HO wave functions in momentum space and the DVR weights
$c_{\mu,l}$, see Ref.~\cite{binder2016}.  The potential corresponding
to the matrix elements is
\ba
V(\mathbf{r}',\mathbf{r}) &=& \sum_{n,n'=0}^N {\psi_{n',0}(r')\over \sqrt{4\pi}}
\langle n' 0|V|n 0\rangle {\psi_{n,0}(r)\over \sqrt{4\pi}} \nonumber\\
&=& {1\over 4\pi} \sum_{\mu=0}^N c_{\mu, 0}\phi_{\mu, 0}(r')\sum_{\nu=0}^N c_{\nu, 0}\phi_{\nu, 0}(r) \ ,
\nonumber\\
\ea
Here, $\psi_{n, l}(r)$ are the Fourier-Bessel transforms of
$\tilde{\psi}_{n, l}(k)$ and we introduced the wave functions
$\phi_{\mu,0}(r)$ that are eigenfunctions of the squared momentum
operator with eigenvalue $k_{\mu, 0}^2$, see
Ref.~\cite{binder2016}. We also used $r=|\mathbf{r}|$,
i.e. $\mathbf{r}=r\hat{\mathbf{r}}$, and
$Y_{00}(\hat{\mathbf{r}})=1/\sqrt{4\pi}$.

We need to compare this with the separable $\delta$ function potential
\ba
W(\mathbf{r}',\mathbf{r}) &=& \delta(\mathbf{r}')\delta(\mathbf{r}) \nonumber\\
&\approx& \sum_{\mu=0}^N{\phi_{\mu, 0}(r')\phi_{\mu, 0}(0)\over 4\pi} \sum_{\nu=0}^N{\phi_{\nu, 0}(r')\phi_{\nu, 0}(0)\over 4\pi} \ , \nonumber\\
\ea
which we re-wrote by using the completeness relation of the wave
functions $\phi_{\mu,0}(r)$ in the finite basis of the DVR.

To relate $V(\mathbf{r}',\mathbf{r})$ to $W(\mathbf{r}',\mathbf{r})$
we compare~\cite{binder2016}
\ba
\phi_{\mu, 0}(r) &=& c_{\mu, 0}\sum_{n=0}^N \tilde{\psi}_{n, 0}(k_{\mu, 0})\psi_{n, 0}(r)
\ea
with the spherical Bessel function
\ba
j_0(kr) = \sqrt{\pi\over 2}\sum_{n=0}^\infty \tilde{\psi}_{n, 0}(k)\psi_{n, 0}(r)
\ea
and find
\be
c_{\mu, 0}\approx \sqrt{\pi\over2}\phi_{\mu, 0}(0) \ .
\ee
Here, the approximate sign enters because the DVR works in a finite HO
basis.  Thus,
\be
V(\mathbf{r}',\mathbf{r}) =
2\pi^2\delta(\mathbf{r}')\delta(\mathbf{r})
\ee
is the contact used in this work. To be precise, our contact is
\ba
V_c &=& 
2\pi^2 g \delta(\mathbf{r})\delta(\mathbf{r}')\tau_-^{(1)}\tau_-^{(2)} \nonumber\\
&=& 2\pi^2 g \delta(\mathbf{r})\delta(\mathbf{r}-\mathbf{r}') \tau_-^{(1)}\tau_-^{(2)} \ , 
\ea
and $g$ is in units of fm$^2$. We varied $g$
between $\pm 1$~fm$^2$ and thus have a strength $|2\pi^2 g|\approx
20$~fm$^2$.

We can now compare to the contact used in
Ref.~\cite{cirigliano2019}. In that paper the contact is written as
\ba
V_S &=& 
-2g_\nu^{\rm NN} \delta(\mathbf{r})\tau_+^{(1)}\tau_+^{(2)} \ ,
\ea
numerical estimates are based on the assumption $g_\nu^{\rm NN}\to
({\cal C}_1 +{\cal C}_2)/2\approx 1$~fm$^2$, see Table~II of
Ref.~\cite{cirigliano2019} for more precise numerical values. Thus,
the contact we used is about a factor of ten larger.

\end{document}